\documentclass[amsmath,amsfonts,aps,prb,preprint,showpacs]{revtex4}

\usepackage{graphicx}
\usepackage{dcolumn}
\usepackage{bm}
\usepackage{color}

\graphicspath{{images/}}
\begin{document}

\title{Size dependent exciton $g$-factor in self-assembled InAs/InP quantum dots}

\author{N. A. J. M. Kleemans\footnote{Electronic mail: n.a.j.m.kleemans@tue.nl}}
\affiliation{Photonics and Semiconductor Nanophysics, COBRA,
Eindhoven University of Technology, P.O. Box 513, NL-5600 MB
Eindhoven, The Netherlands}
\author{J. van Bree}
\affiliation{Photonics and Semiconductor Nanophysics, COBRA,
Eindhoven University of Technology, P.O. Box 513, NL-5600 MB
Eindhoven, The Netherlands}
\author{M. Bozkurt}
\affiliation{Photonics and Semiconductor Nanophysics, COBRA,
Eindhoven University of Technology, P.O. Box 513, NL-5600 MB
Eindhoven, The Netherlands}
\author{P. J. van Veldhoven }
\affiliation{Photonics and Semiconductor Nanophysics, COBRA,
Eindhoven University of Technology, P.O. Box 513, NL-5600 MB
Eindhoven, The Netherlands}
\author{P. A. Nouwens }
\affiliation{Photonics and Semiconductor Nanophysics, COBRA,
Eindhoven University of Technology, P.O. Box 513, NL-5600 MB
Eindhoven, The Netherlands}
\author{R. N\"{o}tzel}
\affiliation{Photonics and Semiconductor Nanophysics, COBRA,
Eindhoven University of Technology, P.O. Box 513, NL-5600 MB
Eindhoven, The Netherlands}
\author{A. Yu. Silov}
\affiliation{Photonics and Semiconductor Nanophysics, COBRA,
Eindhoven University of Technology, P.O. Box 513, NL-5600 MB
Eindhoven, The Netherlands}
\author{M. E. Flatt\'{e}}
\affiliation{Department of Physics and Astronomy and Optical Science
and Technology Center, University of Iowa, Iowa City, Iowa 52242,
USA}
\author{P. M. Koenraad}
\affiliation{Photonics and Semiconductor Nanophysics, COBRA,
Eindhoven University of Technology, P.O. Box 513, NL-5600 MB
Eindhoven, The Netherlands}


\begin{abstract}
We have studied the size dependence of the exciton $g$-factor in
self-assembled InAs/InP quantum dots. Photoluminescence measurements
on a large ensemble of these dots indicate a multimodal height
distribution. Cross-sectional Scanning Tunneling Microscopy
measurements have been performed and support the interpretation of
the macro photoluminescence spectra. More than 160 individual
quantum dots have systematically been investigated by analyzing
single dot magnetoluminescence between 1200\,nm and 1600\,nm. We
demonstrate a strong dependence of the exciton $g$-factor on the
height and diameter of the quantum dots, which eventually gives rise
to a sign change of the $g$-factor. The observed correlation between
exciton $g$-factor and the size of the dots is in good agreement
with calculations. Moreover, we find a size dependent anisotropy
splitting of the exciton emission in zero magnetic field.
\end{abstract}

\date{\today}
\maketitle

\vskip -0.5cm

\section{Introduction}
Self-assembled quantum dots are one of the most promising candidates
to be used as building blocks in quantum information processing.
\cite{Stevenson_Nature, Loss_PRA, Wolf_Science, Doty_PRL} For
instance, single-qubit operations have been proposed by changing the
local effective Zeeman interaction in a quantum dot.
\cite{Kane_Nature, Vrijen_PRA} Control over the exciton $g$-factor
($g_{ex}$), defined by Eq. \ref{g_ex}, is thus highly desirable for
the realization of individual qubits. \cite{Klauser} Moreover, a
sign change of the exciton $g$-factor is very desirable in quantum
information processing and thus there is a strong interest in
quantum dots having a zero $g$-factor due to the structure of the
dot. To investigate the size, shape and composition dependence of
the electron, hole and exciton $g$-factor, theoretical
investigations using the \textbf{k$\cdot$p} approximation
\cite{Nakaoka_PRB2004, Nakaoka_PRB2005, Flatte_PRL} as well as tight
binding calculations \cite{Sheng_PRB, Sheng_PhysicaE} have been
performed on InAs/GaAs dots. The self-assembly process of quantum
dots gives rise to a distribution in size, shape and composition of
the dots and therefore leads to a dot to dot variation of $g_{ex}$.
This opens the possibility to utilize the growth conditions to
engineer $g_{ex}$. \cite{MedeiroRibeiro} However, up to now
experiments on InAs/GaAs QDs did not reveal a strong correlation
between emission energy and $g_{ex}$. \cite{Nakaoka_PRB2005,
Bayer_PRB}

We have performed photoluminescence (PL) measurements on a large
number of single InAs/InP quantum dots in order to investigate the
energy dependence of $g_{ex}$ and its dependence on the structural
properties of individual quantum dots. In this paper we will
demonstrate strong correlations between $g_{ex}$, the diamagnetic
shift and the emission energy of the InAs/InP quantum dots.
Eventually, the size dependence of $g_{ex}$ will lead to a sign
change of $g_{ex}$. The observed correlations can be explained well
by the theoretical trends discussed in Ref. \onlinecite{Flatte_PRL}.
Furthermore, we will analyse the anisotropy splitting of these dots
and correlate this to the height and lateral size of the dots. As
the PL of the InAs/InP quantum dots is tunable to 1.55\,$\mu$m
\cite{Gong_APL, Anantathanasarn_JAP, Poole_JVSTB,
Chauvin_Nanotechnology}, our results show that $g$-factor
engineering is also feasible at telecommunication wavelengths.

\section{Sample growth and characterization}
\subsection{Growth}
Our quantum dots are grown by Metal-Organic Vapor-Phase Epitaxy
(MOVPE). A layer of 100\,nm of InP has been grown on a $n$-doped InP
(100) substrate with a two degrees miscut towards the [110]
direction. Two monolayers (ML) of GaAs were deposited as an
interlayer, thereby reducing the As/P exchange reaction. On top of
this interlayer a 2 ML InAs layer is grown, resulting in the
formation of quantum dots. The quantum dot layer is capped by
200\,nm of InP. For Atomic Force Microscopy (AFM) a layer of surface
quantum dots was grown under the same conditions. From the AFM
measurements we find an average height of the dots of (2$\pm$1)\,nm
and a dot diameter of (34$\pm$5)\,nm. More details about the growth
of these wavelength-tunable InAs quantum dots in InP can be found in
Ref. \onlinecite{Anantathanasarn_JAP}.

\subsection{Macro photoluminescence}
\begin{figure}
\includegraphics[width=8.75cm]{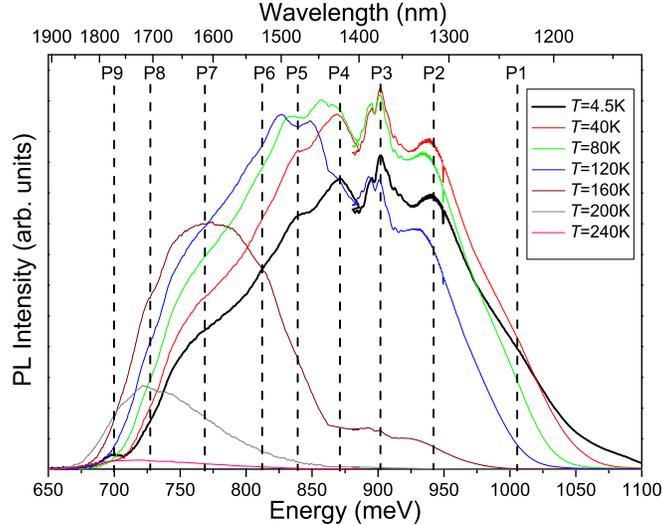}
\caption{\label{figure1} (Color online) PL spectra of a large
ensemble of quantum dots measured at different temperatures. A
multiple peak structure is observed consisting of 9 peaks. The peak
positions at $T=4.5$\,K are indicated by the dotted lines. We
attribute the multiple peak structure to the multimodal height
distribution of the dots. Quantum dots having the smallest height
have luminescence around peak P1.}
\end{figure}
The sample is characterized by temperature dependent PL measurements
performed on a large ensemble of dots. The quantum dots are excited
by a laser operating at 532\,nm and with a spot size of
$\sim$4\,mm$^2$. The macro PL is detected by an InGaAs array
detector up to 1550\,nm and with an InSb single channel detector
above 1400\,nm. The spectra taken at different temperatures are
shown in Fig. \ref{figure1}. These spectra are plotted by matching
at 1450\,nm the spectra obtained by both detectors. Instead of a
single Gaussian distribution, characteristic for highly homogeneous
quantum dots, a series of peaks (P1-P9) is observed. The spectrum at
$T=4.5$\,K displays strong similarities with the ones reported in
Refs. \onlinecite{Heitz_PRB} and \onlinecite{Dion_PRB}; the peaks
were identified as quantum dots with discrete height differences of
1 ML and the dots were modeled accordingly. In the same
way we attribute the different peaks to a multimodal height
distribution of our dots. Quantum dots emitting around P9 at the low
energy side of the spectrum have the largest height, whereas dots emitting at
the high energy side have the smallest height. The width of the
peaks is due to the dot to dot variation of the diameter and
composition. The structure present in peak P3 is still not
understood. A redistribution of carriers over the dots having
different heights occurs for increasing temperatures. At elevated
temperatures the excitons in the QDs with smaller height can escape
and diffuse towards the higher dots, where they are captured and
recombine.

\subsection{Cross-sectional Scanning Tunneling Microscopy}
\begin{figure}
\includegraphics[width=8.75cm]{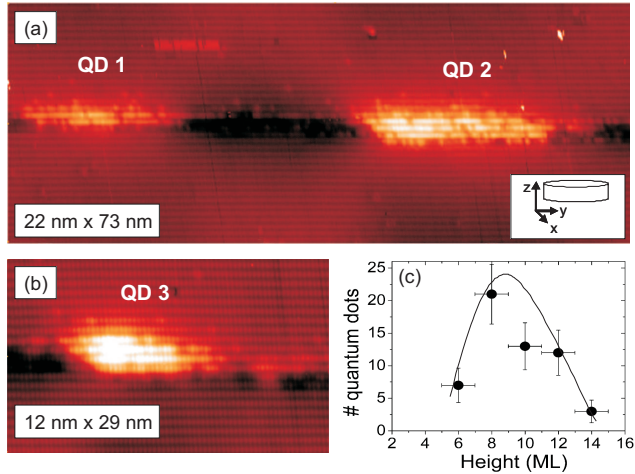}
\caption{\label{figure2} (Color online) X-STM characterization of
InAs/InP quantum dots of (a) 3\,BL (6$\pm$1\,ML), 5\,BL
(10$\pm$1\,ML), and (b) 4\,BL (8$\pm$1\,ML) height. The bright
contrast corresponds to InAs, whereas the dark contrast corresponds
to GaAs. The distribution of the different heights of the dots is
given in (c). The inset in (a) shows the typical disk shape of our
dots.}
\end{figure}
To characterize the dot size, shape and composition we performed
Cross-sectional Scanning Tunneling Microscopy (X-STM), as was done
on similar dots recently. \cite{Ulloa_APL} The measurements have
been performed in constant current mode. Three different quantum
dots are shown in Figs. \ref{figure2}(a) and (b). The images were
obtained at a voltage of $-3$\,V. At these voltages the contrast is
mainly caused by topographic effects due to strain induced surface
relaxation. \cite{Bruls_APL} The bright contrast corresponds to InAs
with the largest lattice constant and the dark contrast is
identified as the GaAs interlayer with the smallest lattice
constant. From these measurements we determine the height of the
dots with bilayer (BL) precision. Note that in X-STM individual ML
cannot be distinguished. Fig. \ref{figure2}(a) shows two different
dots having a height of 3 and 5\,BL and Fig. \ref{figure2}(b) shows
a dot with a height of 4\,BL, which correspond to a height of
(6$\pm$1)\,ML, (10$\pm$1)\,ML and (8$\pm$1)\,ML respectively. For
more than 50 dots the height was measured and the resulting
distribution is shown in Fig. \ref{figure2}(c). The quantum dots
best resemble circular discs, as depicted in the inset of Fig.
\ref{figure2}(a), and therefore the height of the dot is independent
of where the dot is cleaved. The height distribution shows that we
have quantum dots with heights varying between 5 and 15\,ML, which
matches quite well with the 9 peaks we observe in the macro PL. A
height of 5\,ML would then correspond to dots belonging to macro PL
peak P1. Moreover, most dots have a height between 7-9\,ML
corresponding to the part of the PL spectrum which is most intense
(P3-P5).

The X-STM images also show that the lateral sizes of the quantum
dots are less well defined. The largest diameter found by X-STM is
30\,nm and corresponds to the value found by AFM. \cite{Bruls_APL}
The GaAs interlayer is not located between the InAs dot and the InP
substrate, but the InAs dots are rather embedded in the GaAs layer.
Although the GaAs layer suppresses the As/P exchange reaction, the
actual role of this layer in the growth of these dots is still a
matter of further investigation. There appears to be no strong
intermixing of Ga and P inside the quantum dot and therefore we
conclude that our dots consist of almost pure InAs. For all the
studied dots comparable compositions are found. Furthermore, Figs.
\ref{figure2}(a), (b) show that the dot formation preferentially
takes place at the step edges introduced by the miscut of the
substrate.

\section{Magnetoluminescence of individual quantum dots}
\subsection{Experiment}
In order to study the PL of individual quantum dots we use an
aluminium mask on top of the sample, with openings varying between
500\,nm and 1400\,nm. Most measurements have been performed on
openings of 1\,$\mu$m. The excitation is provided by a 635\,nm
wavelength cw laserdiode. We studied quantum dots emitting between
1200\,nm and 1600\,nm using a confocal microscopy setup. The PL was
analyzed in the Faraday configuration in magnetic fields up to 10\,T
aligned parallel with the growth direction. \cite{Bayer_PRB,
Schulhauser_PRB} The polarization is analyzed using an achromatic
quarter wave plate and a linear polarizer. The luminescence was
dispersed by a 75\,cm monochromator and detected by an InGaAs array.
The linewidth varies from dot to dot, and is of the order of
100\,$\mu$eV, limited by the quantum dot linewidth itself. In order
to exclude biexciton luminescence we performed power dependent
measurements and excluded all lines with a superlinear dependence on
the excitation density. \cite{Chavez-Pirson_APL}

\subsection{Correlation between emission energy, exciton $g$-factor and diamagnetic shift}
\begin{figure}
\includegraphics[width=8.75cm]{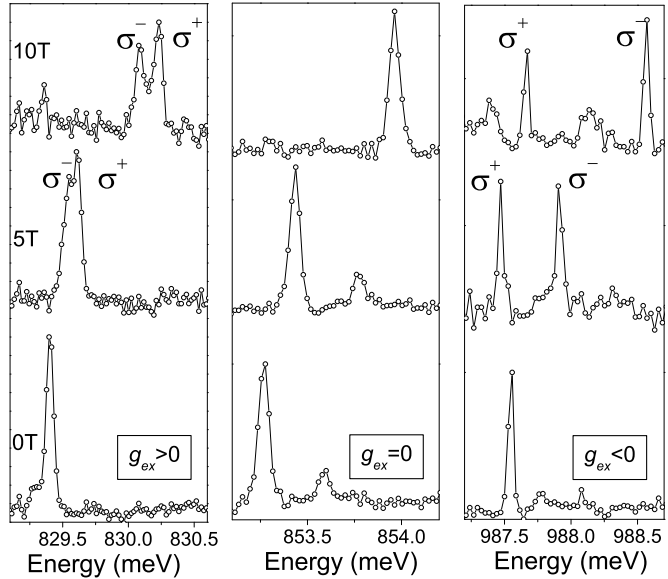}
\caption{\label{figure3} PL of three individual quantum dots showing
from left to right a positive exciton $g$-factor, a quenched
$g$-factor and a negative $g$-factor. The spectra are shown for
magnetic fields of 0\,T, 5\,T and 10\,T in the Faraday
configuration. The polarization was determined with a quarter lambda
plate and a linear polarizer.}
\end{figure}

The emission energy $E(B)$ of an exciton in a quantum dot in a
magnetic field $B$ is in good approximation given by:
\begin{equation}
E(B)=E_0\pm g_{ex}\mu_{B} B+\alpha_{d}B^2\label{EB}
\end{equation}
where $E_0$ is the emission energy at $B=0$\,T,
$\mu_{B}=+5.79\times10^{-5}$eV/T is the Bohr-magneton, and
$\alpha_{d}$ is the diamagnetic coefficient. The second term of Eq.
\ref{EB} is the Zeeman term which gives rise to a spin induced
splitting of the exciton PL in a magnetic field, whereas
$\alpha_{d}$ is linked to the exciton radius. The
magnetoluminescence spectra of three individual quantum dots
emitting at different energies are shown in Fig. \ref{figure3} for
magnetic fields of $B=0$\,T, 5\,T and 10\,T. We observe a clear sign
change of the polarization of the Zeeman splitted lines for the low
energy quantum dot as compared to the high energy dot. Moreover, for
the quantum dot emitting around 850\,meV we observe no Zeeman
splitting at all for magnetic fields up to 10\,T. All three dots
exhibit a diamagnetic shift towards higher energies for increasing
magnetic field.

In order to analyze the data we define $g_{ex}$ as:
\begin{equation}
g_{ex}={\frac{E(\sigma^+)-E(\sigma^-)}{\mu_{B}B}} \label{g_ex}
\end{equation}

\begin{figure}
\includegraphics[width=8.75cm]{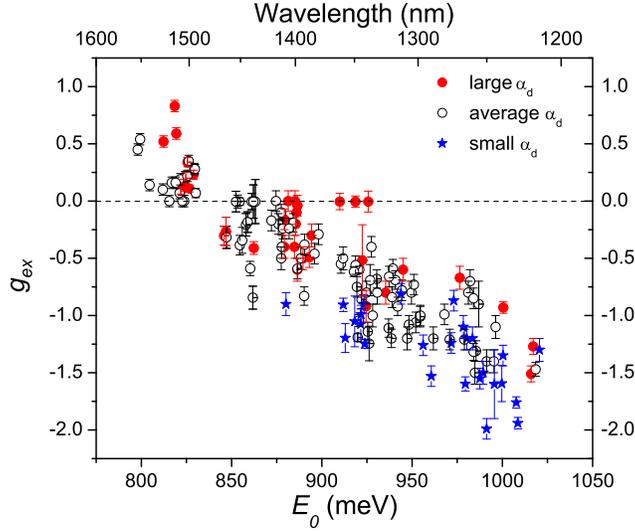}
\caption{\label{figure4} (Color online) The exciton $g$-factor as
function of the emission energy $E_0$ for 164 quantum dots. A sign
change of $g_{ex}$ is observed for dots emitting at low energies.
The quantum dots having a small height have a more negative
$g$-factor as compared to dots having a large height. Moreover dots
having both a small height and a small diamagnetic coefficient
$\alpha_{d}$ (blue stars) , i.e. small lateral size, have the
largest negative $g$-factor. The colors represent different
intervals of $\alpha_{d}$, and correspond to the colors as shown in
the histogram in Fig. \ref{figure5}.}
\end{figure}

Figure \ref{figure3} shows, from left to right, a dot with
$g_{ex}>0$, $g_{ex}\simeq0$ and $g_{ex}<0$. In order to verify the
sign of $g_{ex}$, we also measured control samples with known
$g_{ex}$ in a given direction of the magnetic field and known angle
between the axes of the quarter lambda plate and the linear
polarizer. To reveal the relation between $g_{ex}$ and the emission
energy we investigated the exciton $g$-factor of in total 164
quantum dots. The dependence of $g_{ex}$ on $E_0$ is shown in Fig.
\ref{figure4}. A strong correlation between $E_0$ and $g_{ex}$ is
observed. At large emission energy the exciton $g$-factor changes
its sign and becomes increasingly negative. \cite{Kim_arXiv} The
exciton $g$-factor changes from +0.5 to $-2$ for dots emitting at
775\,meV to 1050\,meV. Since the emission energy of the dot is
mainly determined by the height of the dot, as is inferred from the
macro PL, the dots having a smaller height have a more negative
$g_{ex}$.

\begin{figure}
\includegraphics[width=8.75cm]{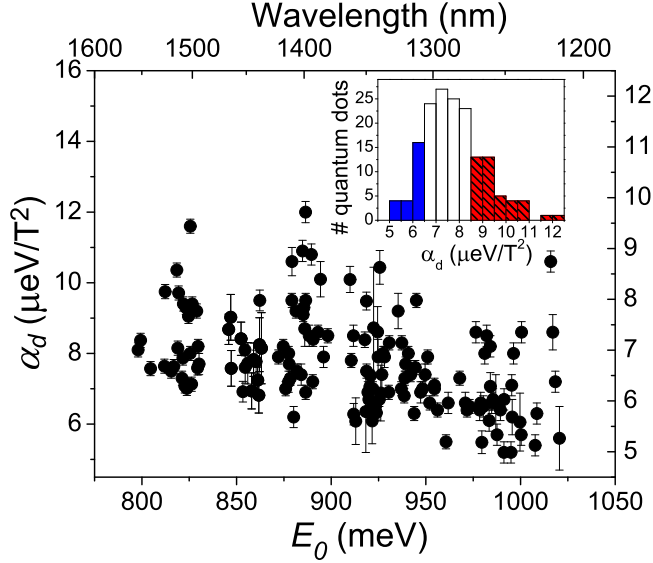}
\caption{\label{figure5} (Color online) The diamagnetic coefficient
as function of the emission energy. There is only a weak correlation
between the diamagnetic coefficient and the emission energy. The
inset shows the histogram of the different values of $\alpha_d$.
Blue corresponds to small values of $\alpha_d$, white to the average
values of $\alpha_d$, and red (hatched) to the large values of
$\alpha_d$.}
\end{figure}

From the magnetic field dependence of the exciton lines we also
extract the diamagnetic coefficient $\alpha_{d}$, which is to good
approximation proportional to the spatial extension of the exciton
wave function, and is therefore a measure for the lateral size of
the dot. \cite{Walck_PRB} To verify that the emission energy is
mainly determined by the height of the dot, we plot $\alpha_{d}$
against the emission energy in Fig. \ref{figure5}. There is only a
weak correlation between emission energy and the diameter of the
dots. We therefore conclude that the change from positive to
negative values of $g_{ex}$ is governed by the quantum dot height.
The weak correlation between $E_0$ and $\alpha_d$ indicates that
dots of smaller (larger) height have on average a smaller (larger)
lateral size. Fig. \ref{figure4} shows that quantum dots emitting at
the same energy have a large variation of the diamagnetic shift. In
order to analyze the importance of the lateral size of the dot on
$g_{ex}$, we specify in Fig. \ref{figure4} three different ranges of
$\alpha_{d}$. These ranges are determined from the distribution of
$\alpha_{d}$ as shown in the inset of Fig. \ref{figure5}, and
correspond to quantum dots with small (blue), average (white) and
large (red) $\alpha_{d}$. We find that dots emitting at the same
energy, but having a smaller lateral size, have a more negative
$g_{ex}$. Thus reducing the size of the dots, i.e. either height or
diameter, will result in more negative values of $g_{ex}$.

\begin{figure}
\includegraphics[width=10cm]{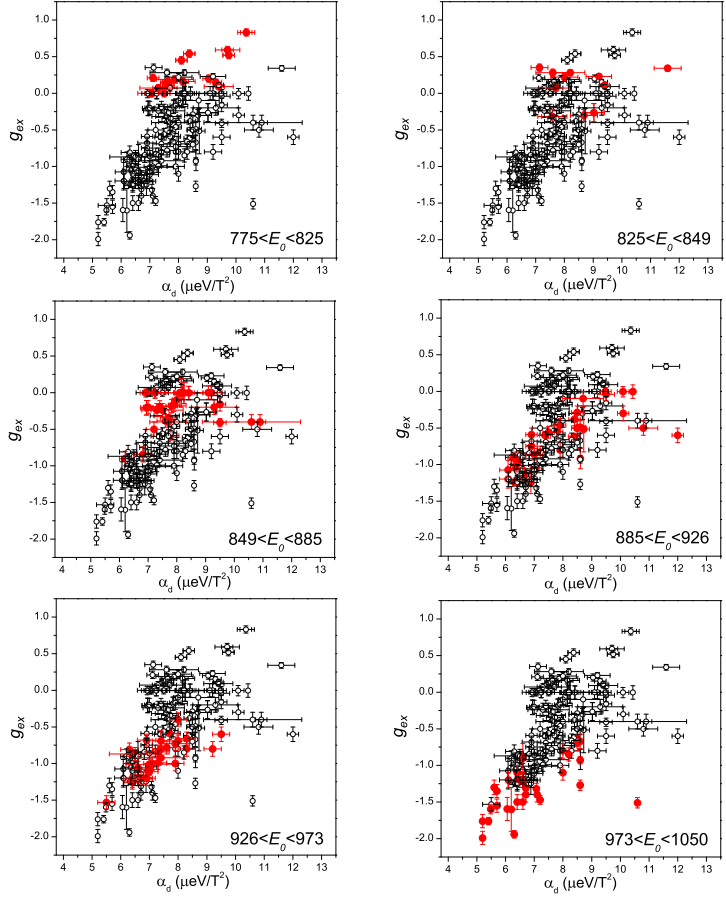}
\caption{\label{figure6} (Color online) The exciton $g$-factor as
function of the diamagnetic coefficient for different emission
energies $E_0$. There is a strong correlation between $\alpha_d$ and
$g_{ex}$. The filled red symbols correspond to the emission range
indicated in separate graphs. These emission intervals correspond to
the discrete peaks P1-P6 in the macro PL spectrum and thus to dots
of different height. The lowest dots which have a smallest lateral
size have the most negative exciton $g$-factor.}
\end{figure}

The relation between $\alpha_d$ and $g_{ex}$ is plotted in Fig.
\ref{figure6}. We find a strong correlation between $\alpha_d$ and
$g_{ex}$. In general there is an increase of the exciton $g$-factor
for increasing $\alpha_d$. The filled red symbols in the different
panels correspond to different emission wavelengths, which
correspond to the energy ranges around the peaks in the macro PL.
The filled symbols in the lower right panel correspond to emission
energies around peak P1 of the macro PL data and correspond to the
dots lowest in height. Quantum dots of the same height, but of
smaller diameter, have a more negative $g_{ex}$. Figure
\ref{figure6} thus shows that quantum dots having the smallest
diameter and height, i.e. the overall smallest size, have the most
negative $g_{ex}$. Increasing the size of the dot results in sign
change of $g_{ex}$, where the dots with the overall largest size
(filled symbols in the upper left panel) have the most positive
$g_{ex}$.

Up to now we assumed that the change in the emission energy of the
dots did not arise from the change in composition of the dots. We
can exclude that composition plays a large role as X-STM did not
show significant compositional variations over the different dots.
Moreover, the influence of the composition on InAs/GaAs dots has
been addressed in several papers, which conclude that there is only
a small effect on $g_{ex}$. \cite{Nakaoka_PRB2004}

To understand the trend towards more negative values of the $g_{ex}$
for smaller dots, we compare our data with the calculations of the
electron ($g_e$) and hole ($g_h$) $g$-factor for InAs/GaAs quantum
dots. \cite{Flatte_PRL} The calculations give $g_e$ and $g_h$ as
function of increasing emission energy. As the composition is fixed
in the calculations, the increase of $E_0$ is only due to the
decreasing size of the quantum dot. The results show that whereas
$g_e$ is relatively insensitive for change in the overall size of
the dot, there is a strong dependence for $g_h$ on the height and
lateral size of the quantum dot. The exciton $g$-factor is defined
by $g_{ex}=-g_e+g_h$. For increasing $E_0$ (i.e. decreasing size of
the dot) the value of $g_e$ increases and $g_h$ decreases, and
therefore they both contribute to a more negative $g_{ex}$. This is
in perfect agreement with our experimental observations. Preliminary
calculations of the $g$-factors for InAs/InP quantum dots indicate
the same trends as for InAs/GaAs quantum dots, although the overall
magnitude of the $g_h$'s is smaller. \cite{PPF-unp} The reduced
strain in InAs/InP quantum dots relative to InAs/GaAs quantum dots
reduces the splitting of the heavy-hole and light-hole band edges in
the dot, and thus there is more light-hole character in the
highest-energy hole state of an InAs/InP quantum dot than in that of
an InAs/GaAs quantum dot. As the light-hole $g$-factor is less
negative than the heavy-hole $g$-factor, this effect leads to less
negative $g_h$'s in InAs/InP dots than in InAs/GaAs dots. Smaller
dots also have smaller light-hole character in the highest-energy
hole state, due to the differing effects of confinement on the heavy
and light hole energies, and thus smaller quantum dots have more
negative $g_h$'s than larger quantum dots, as seen in the measured
$g_{ex}$ trend.

\subsection{Anisotropy splitting}
\begin{figure}
\includegraphics[width=8.75cm]{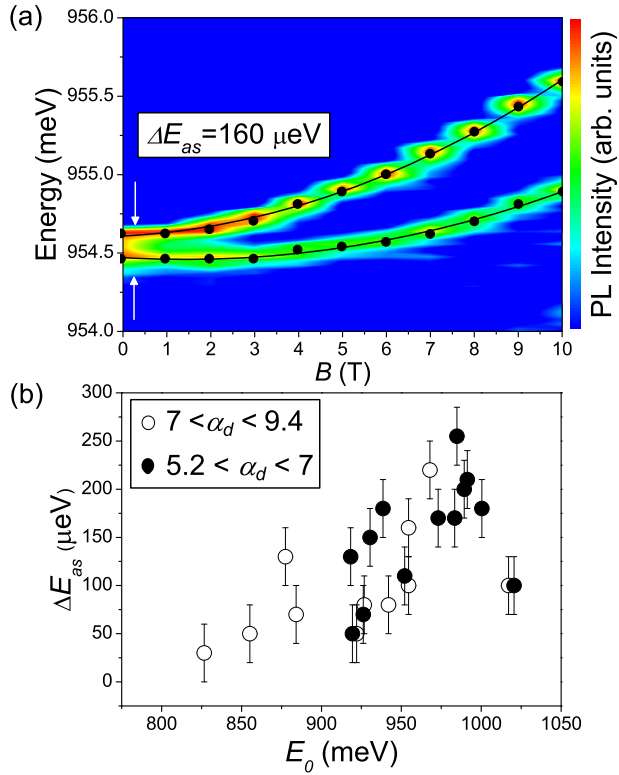}
\caption{\label{figure7} (Color online) (a) Contour plot of the
magnetoluminescence of a dot showing an anisotropy splitting of
$\Delta E_{as}=160$\,$\mu$eV at $B=0$\,T. The blue (white) color
corresponds to low (high) PL intensity. The peak positions used in
the fitting procedure are indicated with the circles and are fitted
by the lines using Eq. \ref{EB}. For this particular dot
$g_{ex}=(-1.00\pm0.09)$ and $\alpha_d=(7.1\pm0.2)$\,$\mu$eV/T$^{2}$.
(b) The anisotropy splitting $\Delta E_{as}$ of in total 24 quantum
dots as function of their emission energy $E_{0}$. The filled
(empty) circles indicate dots having a small (large) diamagnetic
coefficient.}
\end{figure}

Analysis of the single dot spectra showed anisotropy splittings
($\Delta E_{as}$) for 24 quantum dots with a magnitude up to
250\,$\mu$eV in zero magnetic field. The measured values of $\Delta
E_{as}$ are comparable with those found for InAs/GaAs quantum dots.
\cite{Bayer_PRB} As an example a contour plot of the
magnetoluminescence of a quantum dot with $\Delta
E_{as}=160$\,$\mu$eV is shown in Fig. \ref{figure7}(a). Recently,
there has been discussion about the origin of this splitting
\cite{Abbarchi_PRB}, but it is generally believed to arise from the
asymmetry of the footprint of the dot. \cite{Urbaszek_PRL,
Seguin_PRL, Finley_PRB} To demonstrate the dependence of $\Delta
E_{as}$ on the quantum dot size, we plot $\Delta E_{as}$ as a
function of the emission energy in Fig. \ref{figure7}(b). In this
analysis we only treat the subset of quantum dots that exhibit an
anisotropy splitting resolved in our experiments. As shown in Fig.
\ref{figure7}(b), dots having a smaller height, i.e. larger $E_0$,
have in general a larger anisotropy splitting. We believe this is
due to the fact that for quantum dots of lower height the exciton
wave functions are more squeezed in the lateral directions.
Therefore they are more sensitive to the asymmetry of the footprint
of the dot, resulting in larger values of $\Delta E_{as}$.
Nevertheless, higher dots are still sensitive to the confinement
potential asymmetries when they have a large lateral size. This is
depicted in Fig. \ref{figure7}(b) by making a distinction between
dots which have a small and large $\alpha_d$. The anisotropy
splitting for the higher dots is only observed for dots having a
large diamagnetic coefficient ($\alpha_d>7$\,$\mu$eV/T$^2$). In
general we find that both small and large lateral sizes give rise to
an anisotropy splitting for quantum dots of lower height. It should
be noticed that the anisotropy splitting does not occur for the
negatively and positively charged exciton, which supports our
assumption that we are considering the neutral exciton.

\section{Conclusions}
Macro PL and X-STM measurements showed that the studied InAs/InP
dots have a multimodal height distribution. Single quantum dot
luminescence, carried out on a large number of dots, showed a strong
correlation between exciton $g$-factor, diamagnetic coefficient and
emission energy. The strong dependence of $g_{ex}$ on the emission
energy results in a sign change of the exciton $g$-factor. The
trend in $g_{ex}$ is mainly governed by the height variation. We
also demonstrated that the value of $g_{ex}$ is correlated with the
diamagnetic coefficient and conclude that dots with a large diameter
have a smaller $g_{ex}$. In general dots having a smaller overall
size will have a more negative $g_{ex}$ as compared to quantum dots
of larger overall size, which is in agreement with calculations
performed in Ref. \onlinecite{Flatte_PRL}. We also showed that for
several quantum dots the exciton $g$-factor is quenched. This opens
the possibility to evenly tune the sign of $g_{ex}$ by using for
instance electric fields.

We observed anisotropy splittings for InAs/InP quantum dots, and
demonstrated that low dots can give rise to a larger anisotropic
splitting. We conclude that quantum dots with large height and small
lateral size are the most suitable candidates to be used as an
entangled photon source, since this application relies on dots
having small anisotropy splittings. \cite{Stevenson_Nature}

Our study gives a detailed insight into the exciton $g$-factor in
quantum dots and opens a possibility of engineering and controlling
the $g$-factor in individual quantum dots.

\begin{acknowledgments}
The authors like to thank T. E. J. Campbell Ricketts for fruitful
discussions. This work is part of the research program of NanoNed
and FOM, which are financially supported by the NWO (The
Netherlands). M.E.F. also acknowledges support from an ONR MURI.

\end{acknowledgments}

\end{document}